\documentclass[aps,pre,twocolumn]{revtex4-1}

\usepackage{amsmath}
\usepackage{amssymb}
\usepackage{color}
\usepackage{hyperref}
\usepackage{bm}
\usepackage{footmisc}

\usepackage{graphicx}
\usepackage{changes}

\DeclareMathAlphabet{\mathitbf}{OML}{cmm}{b}{it}

\newcommand{\dbar}{{\,\mathchar'26\mkern-12mu d}}

\newcommand{\taubar}{\tau\kern-0.5em\raise0.15ex\hbox{\tiny $\circ$}}

\begin{document}
\title{Finite-size effects on sound damping in stable computer glasses}
\author{Corrado Rainone${}^{1}$, Avraham Moriel${}^{2}$, Geert Kapteijns${}^{1}$, Eran Bouchbinder${}^{2}$, and Edan Lerner${}^{1}$}
\affiliation{${}^1$Institute for Theoretical Physics, University of Amsterdam, Science Park 904, 1098 XH Amsterdam, The Netherlands \\ ${}^2$Chemical and Biological Physics Department, Weizmann Institute of Science, Rehovot 7610001, Israel}

\maketitle

It is well known that disorder gives rise to attenuation of low frequency elastic waves in amorphous solids, even if dynamics are strictly confined to the harmonic regime~\cite{Schober_1992,Schirmacher_PRB_2010,Marruzzo2013,Gelin2016,ikeda_scattering_PRE_2018}. In a recent preprint~\cite{new_scenario_arXiv}, Szamel and coworkers present results regarding sound attenuation rates measured within the harmonic regime in very stable three-dimensional (3D) computer glasses; the key result announced in~\cite{new_scenario_arXiv} is the discovery of a new scenario of sound damping in which the shear (transverse) wave attenuation rate follows $\Gamma_T\!\sim\!k^2$ in the low wavenumber $k$ regime. This scenario has been asserted to be experimentally relevant.

\begin{figure}[!ht]
\centering
\includegraphics[width = 0.50\textwidth]{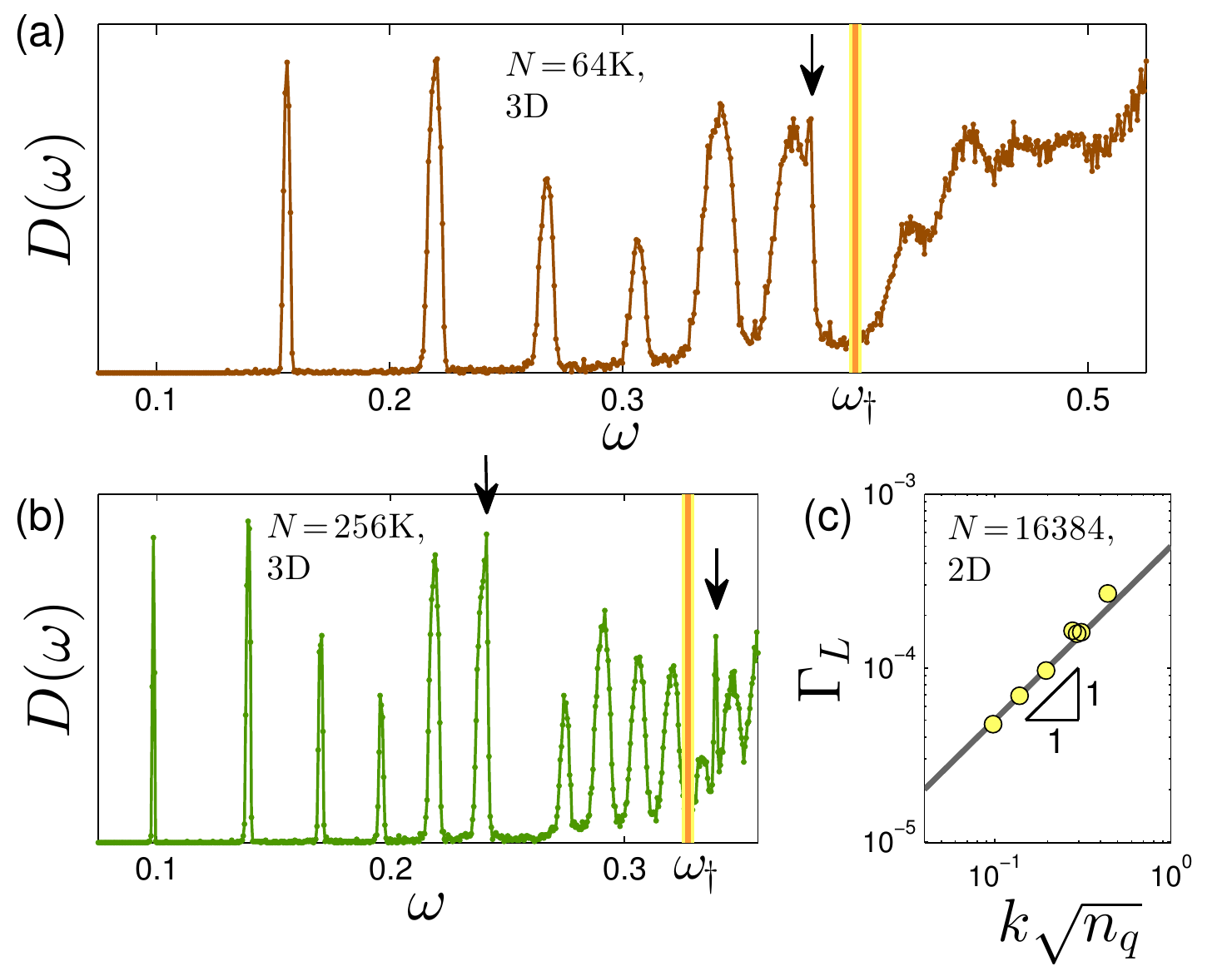}
\caption{The vibrational density of states of 3D stable computer glasses \cite{footnote1} of (a) $N\!=\!64$K and (b) $N\!=\!256$K plotted against frequency \cite{footnote4}. The vertical lines mark our estimation of $\omega_\dagger$, above which phonon bands start to overlap and merge. Arrows mark sound waves. (c) Sound wave attenuation rate $\Gamma_L$ plotted against wavenumber \cite{footnote4} scaled by $\sqrt{n_q}$ (cf.~Eq.~(\ref{eq:NJP})), measured as described in \cite{Gelin2016} for the 7 lowest frequency sound waves in stable 2D computer glasses \cite{footnote5}. }
\label{fig1_note}
\end{figure}

In this brief note we show that the key result of~\cite{new_scenario_arXiv} is in fact a finite-size effect, which has been recently discussed and fully explained theoretically in~\cite{phonon_widths_njp}. Consequently, the sound damping scenario reported on in~\cite{new_scenario_arXiv} is neither new nor experimentally relevant; it will disappear in the thermodynamic limit of macroscopic glasses. Central to understanding the observations of~\cite{new_scenario_arXiv} is the following theoretical prediction for $\Gamma_T$ and $\Gamma_L$ (the sound/longitudinal wave attenuation rate) presented in~\cite{phonon_widths_njp}
\begin{equation}
\label{eq:NJP}
\Gamma_{T,L} \sim \Delta\omega(k) \sim \frac{k\sqrt{n_q(k)}}{\sqrt{N}}\qquad\hbox{for}\qquad k<\frac{\omega_\dagger(N)}{c_T} \ .
\end{equation}
Here $\omega_\dagger$ is a crossover frequency~\cite{phonon_widths_njp}, divided by the shear (transverse) wave-speed $c_T$, below which phonons cluster into discrete bands of disorder-induced width $\Delta\omega(k)$ and degeneracy $n_q(k)$~\footnote{$n_q$ is the number of different solutions to the integer sum of squares problem $q\!=\!n_x^2\!+n_y^2\!+n_z^2$, and $k\!=\!2\pi\sqrt{q}/L$} in finite-size systems composed of $N$ particles, as demonstrated in Fig.~\ref{fig1_note}. Finally, note that it has been shown that $\omega_\dagger(N)\!\sim\! N^{-2/(2\dbar+\dbar^2)}$ in $\dbar$ dimensions~\cite{phonon_widths_njp}.
%%%%%%%%%%%%%%%%%%%%%%%%%%%%%%%%%%%%%%%%%%%%%
%%%%%%%%%%%%%%%%%%%%%%%%%%%%%%%%%%%%%%%%%%%%%%%%%%%%%%

\begin{figure}[!ht]
\centering
\includegraphics[width = 0.50\textwidth]{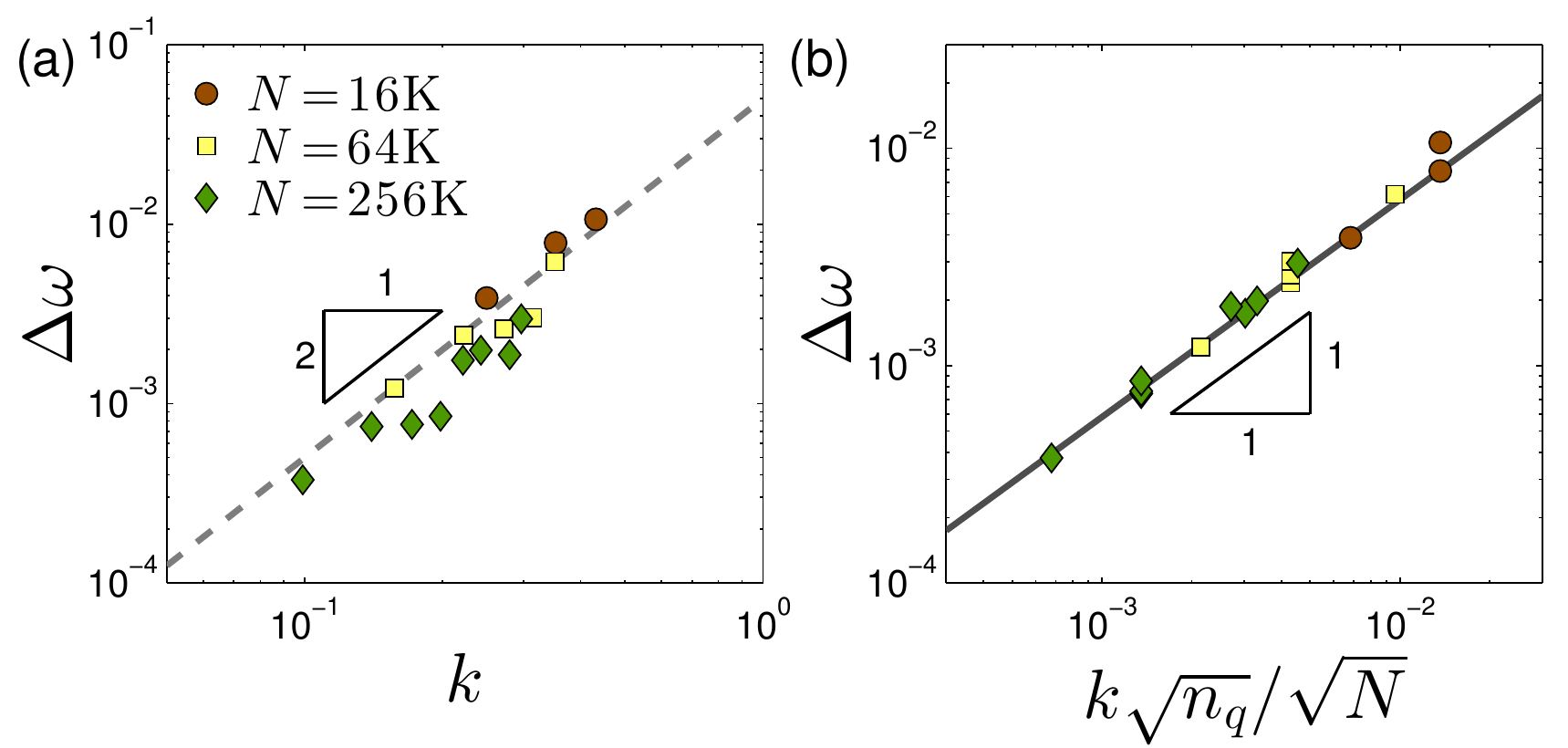}
\caption{(a) Phonon band widths $\Delta\omega$ measured in stable computer glasses, plotted against wavenumber $k$ \cite{footnote4}. An apparent $k^2$ scaling is observed. (b) Same as (a), but plotted against the rescaled wavenumber $k\sqrt{n_q}/\sqrt{N}$, see text for discussion.}
\label{fig2_note}
\end{figure}

To support our main assertion that the low wavenumber regime reported on in~\cite{new_scenario_arXiv} is described by the finite-size theory prediction in Eq.~\eqref{eq:NJP}, we present in Fig.~\ref{fig2_note}a measurements of phonon band frequency widths $\Delta\omega(k)$, extracted (as described in~\cite{phonon_widths_njp}) from the vibrational modes of between $50\!-\!100$ independent stable computer glasses~\cite{footnote1} in $\dbar\!=\!3$, for $N\!=16\hbox{K}, 64\hbox{K}, 256\hbox{K}$, and plotted against the phonon band wavenumber $k$. We find that the widths approximately follow $\Delta\omega(k)\!\sim\! k^2$ (dashed line in Fig.~\ref{fig2_note}a), which corresponds to the observations in~\cite{new_scenario_arXiv}. This scaling, however, is an apparent one; the data in fact follow the finite-size theory prediction in Eq.~\eqref{eq:NJP}, as shown next.
%%%%%%%%%%%%%%%%%%%%%%%%%%%%%%%%%%%%%%%%%%%%%
%%%%%%%%%%%%%%%%%%%%%%%%%%%%%%%%%%%%%%%%%%%%%%%%%%%%%%

In Fig.~\ref{fig2_note}b we plot the phonon band widths against the rescaled wavenumber $k\sqrt{n_q(k)}/\sqrt{N}$. These data establish that the phonon band widths follow the finite-size scaling as given by Eq.~\eqref{eq:NJP}, even for very stable glasses, and consequently that the $k^2$ scaling reported on in~\cite{new_scenario_arXiv} will not persist in the thermodynamic limit $N\!\to\!\infty$, for which $\omega_\dagger(N)\!\to\!0$. Note, though, that the crossover wavenumber $k_\dagger\!\equiv\!\omega_\dagger/c_T$ depends very weakly on system size ($\sim\! N^{-2/15}$ in $\dbar\!=\!3$, consistent with our estimations shown in Fig.~\ref{fig1_note}); varying $N$ by a factor of $4$, as done in~\cite{new_scenario_arXiv}, changes $k_\dagger$ by merely $20\%$, hence explaining the apparent $N$-independent crossover to the finite-size regime observed in Fig.~2b of~\cite{new_scenario_arXiv}.

Finally, an approximate quartic scaling $\Gamma_L\!\sim\! k^4$ of the sound (longitudinal) wave attenuation rate is also reported on in~\cite{new_scenario_arXiv}, exhibiting no finite-size effects as predicted in Eq.~\eqref{eq:NJP}. We assert that for the system sizes employed in~\cite{new_scenario_arXiv}, (all but the very lowest-$k$) sound waves reside above $\omega_\dagger$ (see Fig.~\ref{fig1_note}a-b, where sound waves are marked with arrows), explaining why sound attenuation rates are devoid of finite-size effects in this case. In very stable 2D glasses several sound waves do reside below $\omega_\dagger$, and consequently attenuation rates also follow the predicted finite-size scaling Eq.~(\ref{eq:NJP}), as demonstrated in Fig.~\ref{fig1_note}c.


\begin{thebibliography}{13}
\bibitem{footnote4}
 {Frequencies $\omega$ are expressed in units of $c_T/a_0$, where $c_T$ is the shear
  (transverse) wave speed, $a_0\!\equiv\!(V/N)^{1/3}$ is a typical
  inter-particle distance, and $V$ is the volume; wavenumbers $k$ are expressed
  in units of $a_0^{-1}$; $\Gamma_L$ is expressed in units of $c_L/a_0$, where
  $c_L$ is the sound wave speed.}
\bibitem{footnote5}
{Our 2D computer glasses are of the
  same model and made with the same algorithm as described in~\cite{footnote1};
  they were equilibrated at roughly a quarter of the crossover temperature
  before being instantaneously quenched to zero temperature.}
\end{thebibliography}
\end{document}